\documentclass[prl,showpacs,twocolumn,aps,a4paper]{revtex4}
\usepackage{dcolumn}
\usepackage{amsmath}
\usepackage{graphicx}
\usepackage{latexsym}
\usepackage{amsfonts}
\usepackage{amssymb}
\DeclareGraphicsExtensions{.pdf,.gif,.jpg}

\newcommand{\be}{\begin{equation}}
\newcommand{\ee}{\end{equation}}
\newcommand{\beq}{\begin{eqnarray}}
\newcommand{\eeq}{\end{eqnarray}}

\begin{document}
    
\def\gC{\mbox{\boldmath $C$}}
\def\gZ{\mbox{\boldmath $Z$}}
\def\gR{\mbox{\boldmath $R$}}
\def\gN{\mbox{\boldmath $N$}}
\def\bcalS{\mbox{\boldmath ${\cal S}$}}
\def\bS{\mbox{\boldmath $\Sigma$}}
\def\bG{\mbox{\boldmath $\Gamma$}}
\def\bK{\mbox{\boldmath $K$}}
\def\bH{\mbox{\boldmath $H$}}
\def\bbS{\mbox{\boldmath $S$}}
\def\bM{\mbox{\boldmath $M$}}
\def\bT{\mbox{\boldmath $T$}}
\def\bcalG{\mbox{\boldmath $G$}}
\def\bU{\mbox{\boldmath $U$}}
\def\bQ{\mbox{\boldmath $Q$}}
\def\bV{\mbox{\boldmath $V$}}
\def\bg{\mbox{\boldmath $g$}}
\def\bh{\mbox{\boldmath $h$}}
\def\bq{\mbox{\boldmath $q$}}
\def\bp{\mbox{\boldmath $p$}}
\def\bj{\mbox{\boldmath $j$}}
\def\br{\mbox{\boldmath $r$}}
\def\cZ{{\cal Z}}
\def\ua{\uparrow}
\def\da{\downarrow}
\def\a{\alpha}
\def\b{\beta}
\def\g{\gamma}
\def\G{\Gamma}
\def\d{\delta}
\def\D{\Delta}
\def\e{\epsilon}
\def\ve{\varepsilon}
\def\z{\zeta}
\def\h{\eta}
\def\th{\theta}
\def\k{\kappa}
\def\l{\lambda}
\def\L{\Lambda}
\def\m{\mu}
\def\n{\nu}
\def\x{\xi}
\def\X{\Xi}
\def\p{\pi}
\def\P{\Pi}
\def\r{\rho}
\def\s{\sigma}
\def\S{\Sigma}
\def\t{\tau}
\def\f{\phi}
\def\vf{\varphi}
\def\F{\Phi}
\def\c{\chi}
\def\w{\omega}
\def\W{\Omega}
\def\Q{\Psi}
\def\q{\psi}
\def\de{\partial}
\def\inf{\infty}
\def\ra{\rightarrow}
\def\bra{\langle}
\def\ket{\rangle}
\def\rd{{\rm d}}

\title{Classical Nuclear Motion in Quantum Transport}

\author{Claudio Verdozzi}
\email{cv@teorfys.lu.se}
\affiliation{Solid State Theory,  Lund University, S\"olvegatan 14 A, 223 62 Lund, Sweden}
\author{Gianluca Stefanucci}
\affiliation{Solid State Theory,  Lund University, S\"olvegatan 14 A, 223 62 Lund, Sweden}
\author{Carl-Olof Almbladh}
\affiliation{Solid State Theory,  Lund University, S\"olvegatan 14 A, 223 62 Lund, Sweden}
\date{\today}

\begin{abstract}
An {\em ab initio} quantum-classical mixed scheme for the time evolution 
of electrode-device-electrode systems is introduced 
to study nuclear dynamics in quantum transport.
Two model systems are discussed to illustrate the method.
Our results provide the first example of current-induced molecular desorption
as obtained from a full time-dependent approach
and suggest the use of ac biases as a way to tailor electromigration.
They also show the importance of non-adiabatic effects  
for ultrafast phenomena in nanodevices.
\end{abstract}


\maketitle

The study of electron-nuclei interaction (ENI) in solids has a long and 
important history, dating
back to  the early days of quantum mechanics. Since then, the ENI 
has become a pillar concept
of our understanding in condensed matter \cite{zimanbook}.
Increase in computer power and the introduction of {\it ab-initio}
molecular dynamics \cite{AIMDpaper} have made possible 
quantitative studies of ENI in many materials.

Recent advances in nanotechnology pose new questions about the ENI in 
out-of-equilibrium open systems at the nanoscale \cite{book2}.
Among the techniques used to study the ENI in this context,
quantum transport experiments stand out as a special case, since charge
conduction is at the same time a way to characterize the nanodevice
and a property to be exploited in its operating regime. A case in point is
molecular junctions, where electron injection
can stimulate local vibrations \cite{Science,Nature} and possibly 
electromigration \cite{ys1996}. More generally,
describing ENI in time-dependent (TD) quantum transport is expected to become of great
technological interest,
since  future nanodevices will operate
under the influence of ever faster time varying external fields.
Accordingly, those regarded at present as marginal 
transient effects will 
soon become center stage features. 
Assessing and engineering the ENI is then a key ingredient 
to increase the device efficiency \cite{Hyldgaard,b2005}. 
To date, most theoretical studies have addressed the ENI in
steady-state phenomena \cite{emberly,ths2001,bstmo2003,vpl2002,pfb2005,fbjl2004,bcg2005} 
and often perturbatively in the nuclear displacements 
\cite{pfb2005,fbjl2004,bcg2005}. 
Going beyond the harmonic  approximation and including the ENI in 
a first-principles, TD
framework is a difficult theoretical task  which
has received so far scarce attention \cite{yssb1999, JPCM05cinesi, Horsfield, Todorov2}.

In this work we propose a first principle approach to
TD quantum transport which treats the nuclear motion
in the Ehrenfest dynamics (ED), and the electrons
within Time-Dependent Density Functional Theory (TDDFT)\cite{rg1984}.
The ED has been extensively used in several contexts; 
in quantum transport, it was considered
in \cite{Horsfield}. The ED correctly displays many important features
of ENI, but gives an incomplete account of the Joule heating of the 
nuclei by the electrons \cite{Todorov2}. However, such effect
is not the aim of this first work.  Here we use 
ED which, while treating ENI at the mean field level, 
includes, as opposed to the Born Oppenheimer Approximation,
electronic transitions.
Our approach to transport permits the 
description of transient effects:
it can be used to describe history dependent currents,
hysteresis phenomena, etc.
Another main advantage is that
it can perform the TDDFT-ED of devices
connected to infinitely long leads.

We illustrate the approach with 
two model devices, where the electrons 
interact only via the ENI: a Holstein wire (D1)
and a diatomic molecule (D2). We choose these two rather different
systems to show the versatility of our method
in discussing transient phenomena
and overcoming the limitations of adiabatic treatments.
More in detail, our results show that: 
(i) For weak electron-phonon ($e$-ph) couplings, D1 exhibits an almost periodic 
nuclear displacement (with period=1/density), reminiscent of a Peierls  
distortion. On applying a dc bias, the nuclei oscillate (with decreasing
amplitude) and the period changes to accommodate the current flow.
On increasing the $e$-ph coupling, D1 changes from conducting to insulating.
(ii) D2 is deformed by a small, suddenly switched on dc bias.
Above a critical value of the bias, the molecule dissociates.
This is the single most important result of this work. To our 
knowledge, it provides the first example of current-induced molecular 
desorption as emerging from the full TD dynamics of a nanodevice.
(iii) The desorption cannot be described in
any adiabatic formalism, since it is due to electronic excitations 
induced by the nuclear motion.
(iv) The desorption can be tuned by 
the intensity and frequency of an ac bias, suggesting a way to control 
electromigration in molecular devices.

\noindent {\it The Method}. Following Refs. \cite{c1980,sa2004,ksarg2005}, 
in the initial ground state the central region (C) is in contact to seminfinite left (L)
and right (R)  leads. With ENI, the Hamiltonian for C reads
\be
\hat{H}_{\rm C}[{\bf x}]=\sum_{i,j=1}^{M}V_{ij}({\bf 
x})c^{\dag}_{i}c_{j},
\ee
where $M$ is the number of one-electron states of C and
${\bf x}=(x_{1},\ldots,x_{N})$ are the nuclear coordinates.
Outside C, the nuclei are clamped.
In the Kohn-Sham (KS) scheme of  TDDFT \cite{rg1984},  $V_{ij}$ would be
the $(i,j)$ matrix element of the KS Hamiltonian. 
The nuclear classical Hamiltonian is
$H_{\rm cl}=\sum_{k=1}^{N}p_{k}^{2}/(2m_{k})+U_{\rm cl}({\bf x})$.
The dynamics of the system is governed by
\begin{eqnarray}
i\frac{\rd}{\rd t}|\Q\ket&=&\hat{H}_{\rm el}|\Q\ket,\label{qd}\\
m_{k}\frac{\rd^{2} x_{k}}{\rd t^{2}}&=&
-\frac{\de}{\de x_{k}}\left(U_{\rm cl}+
\bra\Q|\hat{H}_{\rm C}|\Q\ket\right),
\label{cd}
\end{eqnarray}
where $|\Q(t)\ket$ is the many-electron state at time $t$, and
$\hat{H}_{\rm el}[{\bf x}]$ is the electron Hamiltonian 
of the contacted system L+C+R. 
Given a configuration ${\bf x}$, 
$\hat{H}_{\rm el}[{\bf x}]$ is a free-particle Hamiltonian. 
The many-electron ground state $|\Q_{g}[{\bf x}]\ket$ consists of 
bound, resonant, fully reflected waves, plus
left and right going 
scattering states.
The parametric dependence of $\hat{H}_{\rm el}$ on ${\bf x}$ renders 
every eigenstate a function of ${\bf x}$. The ground state value 
${\bf x}={\bf x}_{g}$ is computed using a 
damped ground-state dynamics:  starting from an initial ${\bf x}_{0}$ 
the coordinates are evolved according to 
$m_{k}\ddot{x}_{k}=-\gamma\dot{x}_{k}-\de\left(U_{\rm cl}+
\bra\Q_{g}|\hat{H}_{\rm C}|\Q_{g}\ket\right)/\de x_{k}$, with $\g$ 
the friction coefficient. 

Having ${\bf x}_{g}$ and the corresponding $\{\q\}$ one-electron orbitals, we 
apply an external bias and evolve the system. Assuming metallic electrodes and 
instantaneous screening, the size of C is chosen so that the potential drop 
at any time occurs entirely in C. Thus, the TD part of the electrode
Hamiltonian is a spatially uniform shift $U_{\eta}(t), \eta=$L,R. We use a novel
mixed quantum-classical evolution algorithm, which combines 
a recently proposed generalization of the Crank-Nicholson
method \cite{ksarg2005} for the $\{\q\}$ with a Verlet-like integrator  for the 
${\bf x}$.  Schematically, in terms of the discretized time
$t_{m}=2m \d$,
\begin{equation}
\left\{ 
\begin{array}{lll}       
\left\{\q^{(m+1)}\right\}=\left\{\bbS[t_{m},{\bf 
x}^{(m)}]\q^{(m)}\right\}
\\ \\
\left\{
\begin{array}{l}
p_{k}^{(m+1)}=p_{k}^{(m)}+2\d F_k[{\bf x}^{(m)},\left\{\q^{(m+1)}\right\}]
\\
x_{k}^{(m+2)}=x_{k}^{(m)}+4\d p_{k}^{(m+1)}/m_{k}
\\
p_{k}^{(m+2)}=p_{k}^{(m+1)}+2\d
F_k[{\bf x}^{(m+2)},\left\{\q^{(m+1)}\right\}]
\end{array}
\right.
\\ \\
\left\{\q^{(m+2)}\right\}=\left\{\bbS[t_{m+1},{\bf 
x}^{(m)}]\q^{(m+1)}\right\}
\end{array}
\right.
\label{te}
\end{equation}
with $\q^{(m)}=\q(t_{m})$, ${\bf x}^{(m)}={\bf 
x}(t_{m})$ and ${\bf p}^{(m)}={\bf p}(t_{m})$. The unitary matrix $\bbS$ 
depends on time through ${\bf x}(t)$ and the TD bias
$U_{\eta}(t)$, $\eta=$L,R. Full details 
of the electronic evolution can be found in Ref. \cite{ksarg2005}. 
The force $F[{\bf x},\{\q\}]$ is given by 
the r.h.s. of Eq.(\ref{cd}).

To illustrate the method, we describe electrodes L and R in terms of 
one-dimensional tight-binding (TB) Hamiltonians with a hopping $V$ between 
nearest neighbor sites (TB 
parametrization of Au wires suggest  $|V|$ in the range $ 0.3\div 1.0$  eV).
Left and right going scattering states have 
energy $\ve$ within the band $(-2|V|,2|V|)$ and can be obtained by 
solving the Schr\"odinger equation in C with appropriate boundary 
conditions. 
Bound state eigenenergies $\ve_b<-2|V|$  satisfy 
${\rm Det}\left[\ve_b{\bf 1}-\bH_{\rm C}-\bS(\ve_b)\right]=0$, and 
the associated wavefunction is given in C by the kernel of
$\left[\ve_{b}{\bf 1}-\bH_{\rm C}-\bS(\ve_{b})\right]$
($\bH_{\rm C}$ is the projection of $\hat{H}_{\rm C}$ onto the 
one-electron Hilbert space,  $\bS(\w)$ is
the embedding self-energy). 
The topology of region C 
might also lead to states rigorously confined in C (see below). These are 
resonant eigenstates of the uncontacted $\bH_{\rm C}$ with zero
amplitude at the interface with the two electrodes.\newline
%
%
\noindent {\it Model device D1}. 
The semiclassical Holstein model is a valuable
tool to gain insight into $e$-ph interactions
\cite{km2005,kse1995}.
Here, we investigate TD transport through a 
Holstein wire described by 
\be
\hat{H}_{\rm C}=
V\sum_{i=-M}^{M-1}\left(c^{\dag}_{i}c_{i+1}+{\rm h.c.}\right)
-g \sum_{i=-M}^{M}x_{i}\hat{n}_{i},
\ee
where $\hat{n}_{i}=c^{\dag}_{i}c_{i}$ is the local density operator 
and $x_i$ are the phonon coordinates. 
\begin{figure}[htbp]
\includegraphics*[width=.47\textwidth]{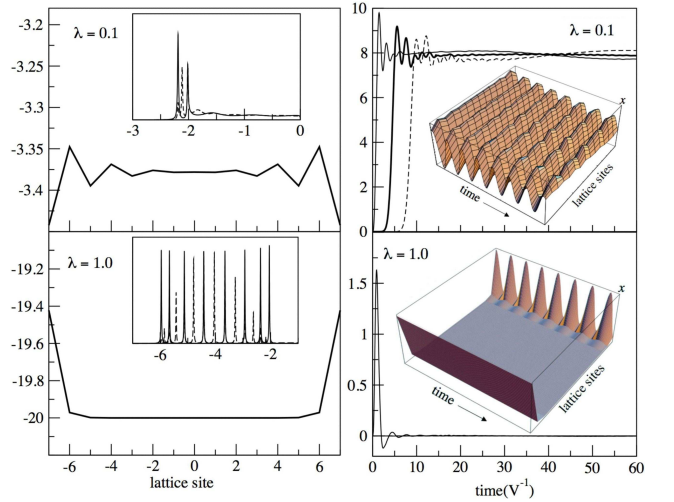}
\caption{Left panel: Ground state displacement  $x_{i}$,  
$i=-7,\ldots,7$, at weak ($\l=0.1$) and strong ($\l=1.0$) coupling. 
The inset shows the LDOS (energy in units of $|V|$) on site $-6$ 
(dashed line) and site $0$ (solid line). Right panel: current $I(t)$
(in units of $10^{-2}|V|$) at weak and strong coupling along the bond 
(-7,-6) (solid thin), (0,1) (solid thick) and $(6,7)$ (dashed)
after the sudden switching on of a bias $U_{\rm L}=0.5 |V|$ in electrode 
L. The insets show a 3D plot of ${\bf x}(t)$ between  $t=0$ and $t=480 |V|^{-1}$ (in the 
top panel the range of ${\bf x}$ is between -3.2 and -4 while in the 
bottom panel is between -19.4 and -20).}
\label{fig1}
\end{figure}
The nuclear classical potential is 
$U_{\rm cl}({\bf x})=\frac{1}{2}\sum_{i=-M}^{M}m_{i}\w^{2}_{0}x^{2}_{i}$.
The strength of the $e$-ph interaction is determined by
the dimensionless parameter $\l=g^{2}/(2V\w_{0})$.
In the original derivation by Holstein, $x_i$ corresponds to an internal coordinate
(bond-length) at the $i$-th site.
We study D1 at half-filling ($\ve_{\rm F}=0$), in the adiabatic regime ($\a=\w_{0}/V=0.1)$
at weak ($\l=0.1$) and strong ($\l=1)$ coupling.
A damped ground-state dynamics, as  described above,
was used to get the converged ground-state $ x_{i}$  (Fig. \ref{fig1}, left)
for a region C with $M=7$.
The energy spectrum 
between $-2|V|$ and $\ve_{\rm F}$ was discretized and good convergence 
was achieved with $N_k=500$ meshpoints. 
For $\lambda=0.1$, a Peierls-like distortion is seen:
an even-odd behavior of $x_{i}$ is manifest (in general, $P=1/n$, and here $n=0.5$),
but exact periodicity is prevented by the finite size of C.
The inset shows the local density of states (LDOS) close
to the interface and in the center, respectively. At 
$\ve <-2|V| $ we observe three peaks due to
three bound states. The picture changes dramatically at $\lambda=1$:  
The number of bound states equals  
the dimension of C and, almost uniformly, $x_i \simeq -20$. 
Only the $x_i$ close to the interfaces are slightly above this value.

In Fig. \ref{fig1}, right, we plot the 
current $I(t)$ in three different points of C after  
suddenly switching on a bias $U_{\rm L}=0.5 |V|$ in the left electrode.
All calculations use a time step $\d=0.01|V|^{-1}$.
For both $\lambda=0.1, 1.0$, the current in the (short) transient is similar to 
the case with the $x_{i}$ clamped (not shown), since electrons 
are much faster than nuclei. At $\l=0.1$ 
the steady current shows superimposed oscillations of 
frequency $\w_{0}$. Instead, for $\l=1$,
all the $2M+1=15$ bound states in C are occupied
(including the Hartree potential would have led to a 
significant reduction of this excess density in C),
no current fluctuations occur in the center and at the right interface
and only a very short transient is observed at the left interface.
The insets display the 
TD Peierls distortion. For $\lambda=0.1$
all the $x$'s oscillate with an exponentially decreasing amplitude 
(within our simulation time $t=480 |V|^{-1}$). The overall shape of the $x_{i}$ changes to 
accommodate the net electron flow. For $\l=1$
no current flows and only the $x_{i}$ close 
to the left interface oscillate. 
The decay of the amplitudes
of the $x_i$ is partially due to
the inefficiency of ED in transferring energy from
electrons to nuclei.
\begin{figure}[t]
\includegraphics*[width=.47\textwidth]{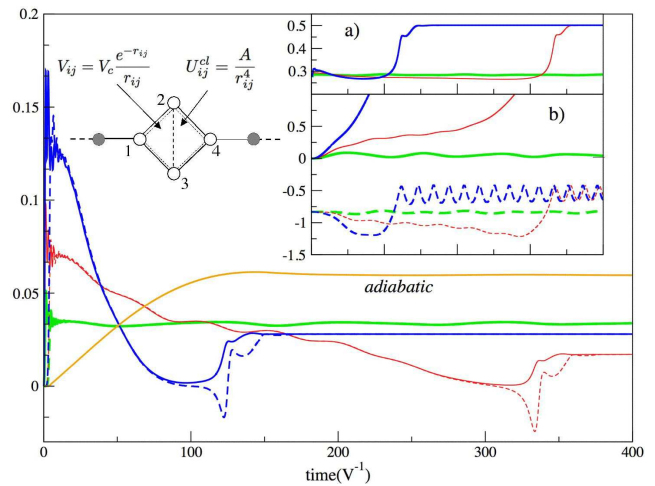}
\caption{$I_{\rm L,R}(t)$ (in units of $|V|$) at the left 
(solid line) and right (dashed line) interface for 
a sudden switching 
$U_{\rm L}=0.25|V|$ (green[light grey]), $0.5|V|$ (red [thin dark grey]) and $1.0|V|$ (blue [thick black]). 
Inset a): Time dependent density $n_{3}$ of nucleus 3 (by symmetry 
$n_{2}=n_{3}$ at any time). Inset b): 
$x_{3}(t)$ (solid line) and $y_{3}(t)$ (dashed line) of nucleus 3 (by 
symmetry $x_{2}=x_{3}$ and $y_{2}=-y_{3}$ at any time). Time scale 
and color coding in insets a) and b) is the same as in the main figure.
A schematic of the device D2 and the adiabatic result (labeled curve)
 are also shown.
}
\label{fig2}
\end{figure}
We also considered the sudden removal of a bound electron, as
obtainable for example by optical means: in this 
case D1 provides a strong transient oscillating response (not shown). On speculative grounds,
such behavior could be used for ENI based photosensors.

\noindent {\it Model device D2}.  We consider a central region C with the simplest 
non-trivial topology, a four-atom ring. This is our model molecular
device D2, with nuclear positions ${\bf r}_{i}\equiv (x_i, y_i),\;i=1,..,4$
(Fig. \ref{fig2}, top-left). In D2, only nuclei 2 and 3 (N2,3) are let to move in the $xy$ plane.
The origin of the $xy$ plane is the midpoint  of  nuclei 1 and 4.
For the hopping parameter we choose the form \cite{gsp1989}
$V_{i\neq j}=V_{c}\frac{e^{- r_{ij}}}{r_{ij}}, V_{i=j}$=0, 
where $r_{ij}=|{\bf r}_i-{\bf r}_j|$.
The purely repulsive classical term is given 
by $U_{\rm cl}=\frac{1}{2}\sum_{i\neq j}A/r_{ij}^{4}$. 
We tune the parameters $V_{c}$, $A$ and $r_{14}$ to ensure a reasonable 
domain of structural stability for D2. For
$V_{c}=4V$, $A=0.75|V|$ and $r_{14}=2$, D2 is stable against deformations 
up to $\sim 10\%$ of the 
equilibrium distances. Also, we get the  ground state positions
${\bf r}_{2}=(0,0.8313)$ and ${\bf r}_{3}=-{\bf r}_{2}$
(the $2-3$  symmetry remains true in the presence of the bias). 
The LDOS has one peak below $-2|V|$, i.e. one bound 
state. There is  also a resonant state 
$|\q\ket=\frac{1}{\sqrt{2}}(|2\ket-|3\ket)$ with energy $|V_{23}|>0$ 
inside the band. The system is taken at half filling, 
and we switch on a bias 
$U_{\rm L}$ at $t=0$.
As for D1, $\d=0.01|V|^{-1}$, $N_k=500$; for the masses  of N2,3 
we choose  $100|V|^{-1}$. 

In Fig. \ref{fig2} we plot the TD
currents $I_{\rm L}$ and $I_{\rm R}$ at the left and right 
interfaces, respectively. At small bias $U_{\rm L}=0.25 |V|$,  $I_{\rm L,R}$
rapidly increase and after a short transient (with the nuclei 
essentially still) start to 
oscillate around a steady value. Inset b) shows the 
corresponding nuclear dynamics. 
The equilibrium rhombic geometry 
changes and the molecule gets deformed in the biased system,
with N2,3 having damped oscillations around two new positions
(for the damping and ED, considerations similar to the case of D1 apply).
We also notice  from inset a) that the charge density of N2,3 slightly increases.

Highly interesting is the strong bias case $U_{\rm L}=1.0|V|$. 
After a very short transient, $I_{\rm L, R}$  sharply decrease (rather than 
oscillating around a steady value) and become zero at 
$t_{0}\simeq 100|V|^{-1}$. After $t_{0}$, $I_{\rm L,R}$ separate:
$I_{\rm L}$ increases while $I_{\rm R}$ decreases, reaches a negative 
minimum and then increases to eventually re-join $I_{\rm L}$ at $t_{1}\sim 160|V|^{-1}$. 
For $t>t_{1}$, $I_{\rm L}\simeq I_{\rm R}$ and their value equal the 
steady current (calculated from the Landauer formula) of the 
chain without N2,3. The behavior of  $I_{\rm L,R}$ can be understood 
looking at the nuclear dynamics (inset  b): the force exerted by the electron flow is strong 
enough for atomic migration to occur. N2,3 are pushed to the 
right by the current, overcome the confining potential and 
get dissociated from region C. 
Thereafter, they form a diatomic molecule vibrating along $y$ (see inset b) 
and traveling along $x$ at uniform speed.
The pronounced minimum of $I_{\rm R}$ shortly 
after $t_{0}$ is due to a sudden charge transfer 
from electrode R (via atom 4) to the diatomic molecule when N2,3 
pass above nucleus 4. This is confirmed in inset a) where the  
density suddenly increases in correspondence of the minimum in 
$I_{\rm R}$. We also note that the total density of the dissociated 
molecule is about 1 (exact charge quantization would occur in 
the adiabatic approximation only). Here, including the Hartree
potential would not lead to qualitative changes of the dissociation
process.  A more realistic model should also
include the possibility for the molecule
to chemisorb onto the electrode, to fragment etc., but
this is beyond the scope of the present Letter. 

To address the dissociation mechanism, we study two
different $U_{\rm L}(t)$, with the same asymptotic value $0.5|V|$.
For a sudden switching, one observes a behavior delayed but similar
to the $1.0|V|$ case above.
Instead, for an adiabatic switching \cite{note} 
the dimer does not dissociate ( in Fig. \ref{fig2}, the current
is displayed in orange and labeled ``adiabatic'' ).
We conclude that electronic excitations induced
by the nuclear motion play a crucial role in the electromigration, 
a role that can not be accounted for by any adiabatic formalism
(without transients the nuclei experience no force).
However, we observe that ED might
overestimate the value of the critical bias for which
dissociation occurs.
%
\begin{figure}[t]
\includegraphics*[width=.47\textwidth]{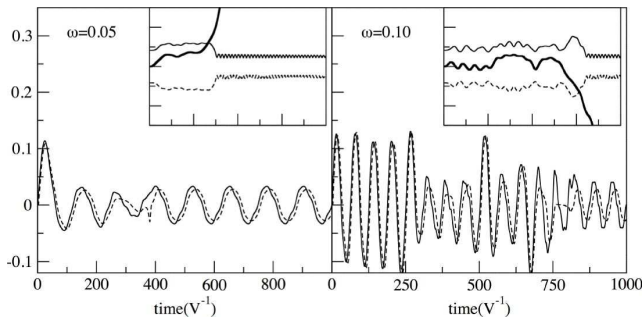}
\caption{Left panel: $I_{\rm L,R}(t)$ (in units of $|V|$) at the left 
(solid line) and right (dashed line) interface for  $U_L=1.0|V|, \omega=0.05|V|$.
In the inset,  $x_2(t)=x_3(t)$ (thick solid) and $y_2(t)$ (thin solid), $y_3(t)$  (dashed).
The time scale is the same as in the main panel.
Right panel: same as left, but with $\omega=0.10|V|$.
Left and right insets have the same vertical scale .}
\label{fig3}
\end{figure}

We next examined the response of D2 when subject to a high amplitude ac bias 
$U_{\rm L}(t)=U_{\rm L}\sin(\omega t)$, with $U_{\rm L}=1.0|V|$.
At low $\omega$ (Fig. \ref{fig3}, left), the system qualitatively behaves as in the dc case,
as clear from both current (main left) and coordinates (inset left) panels.
The nuclei overtake the barrier before the change in $U_{\rm L}(t)$ produces 
a force able to ``recall'' them back. 
After the desorption,  $I_{\rm L,R}$ oscillate as 
they would for a linear chain without N2,3. 
At larger $\omega$ (Fig. \ref{fig3}, right) the dissociation is delayed, since
atoms 2 and 3 are recalled back  a few times before leaving region C. Eventually,
for high enough frequency ($> 0.3|V|$, not shown) atoms 2 and 3 
oscillate without leaving C (within our simulation time).
These results point to a possible
use of ac biases  as a way to tailor  molecular desorption in nanodevices.

In conclusion, we presented a mixed quantum-classical 
scheme to describe electron-nuclei interactions in quantum transport.
The scheme is straightforwardly amenable to an \textit{ab-initio} implementation.
Our results show the necessity to go beyond
adiabatic schemes and to use a full time-dependent, nuclear-dynamics approach 
for ultrafast transient phenomena which are expected to become important 
in future generation nanodevices. 
We acknowledge useful discussions with G. J. Ackland, P. Bokes and 
E. N. Economou. This work was supported by the European Community 
6th framework Network of
Excellence NANOQUANTA (NMP4-CT-2004-500198).


\begin{thebibliography}{10}
\bibitem{note}
We take  $U_{\rm L}(t)=0.5|V| \sin (\pi t/2t_{0}),  t \le t_{0}=150|V|^{-1}$.
To smoothen the switching
as much as possible, we damp the nuclear motion using
a friction $\gamma=0.07$ up to $ t < 200|V|^{-1}$ and let the 
system evolve without friction afterwards.
\end{thebibliography}
\end{document}